\documentclass[letterpaper, 10 pt, conference]{ieeeconf}  
\IEEEoverridecommandlockouts                              
\usepackage{ntheorem}
\usepackage{hyperref}
\usepackage{textcomp}
\let\labelindent\relax
\usepackage{enumitem}
\usepackage{graphicx}
\usepackage{epstopdf}
\usepackage{float}
\usepackage[nocomma]{optidef}
\usepackage{dsfont}
\usepackage{times}
\usepackage{bm}
\usepackage{fancyhdr,graphicx,amsmath,amssymb}
\usepackage[caption=false,font=footnotesize]{subfig}
\usepackage{cite}
\usepackage{color}
\usepackage[algo2e,ruled,vlined]{algorithm2e}
\usepackage{placeins}
\usepackage{cases}

\usepackage{multirow}
\usepackage{makecell}

\newtheorem{theorem}{Theorem}
\newtheorem{remark}{Remark}

\newtheorem{lemma}{Lemma}

\newcommand{\prth}[1]{\left( #1 \right) }
\newcommand{\sqr}[1]{\left[ #1 \right] }

\renewcommand{\H}{\mathbb{H}}

\newcommand{\K}{\mathcal{K}}
\newcommand{\N}{\mathcal{N}}

\newcommand{\R}{\mathbb{R}}

\title{ Incentive Design in Competitive Resource Allocation: \\ Exploiting Valuation Asymmetry in Tullock Contests   }

\author{
    Gilberto D\'iaz-Garc\'ia$^{*}$,
    Keith Paarporn$^{**}$,
    Jason R. Marden$^{*}$
    \thanks{$^{*}$Department of Electrical \& Computer Engineering, University of California, Santa Barbara, Santa Barbara, CA (e-mail: gdiaz-garcia@ucsb.edu, jrmarden@ucsb.edu ) }
    \thanks{$^{**}$Department of Computer Science, University of Colorado, Colorado Springs, CO (e-mail: kpaarpor@uccs.edu) }
    \thanks{ This work is partially supported by AFOSR Grants \#FA9550-21-1-0203 and \#FA9550-25-1-0245FA9550-25-1-0245 and NASA Grant \#103215. }
}

\begin{document}

\maketitle

\begin{abstract} 
    In competitive resource allocation, a central coordinator may seek to gain an advantage not by directly controlling subordinate agents, but by strategically manipulating the information they receive. We study this problem within the framework of multi-player Tullock contests, where the coordinator influences subordinate players by designing their reported valuations of the contested prize, a mechanism that preserves the Tullock structure of the subordinates' objectives and thereby enables tractable equilibrium analysis. We first characterize the Nash equilibrium of the general multi-player Tullock contest, establishing how valuations and per-unit costs jointly determine equilibrium bids and payoffs. We then derive the optimal reported valuations for a coordinator managing two subordinates against a single opponent, and show that the structure of the optimal solution extends to contests with an arbitrary number of subordinates, reducing the coordinator's optimization to a two-variable problem regardless of system size.

\end{abstract}

\IEEEpeerreviewmaketitle


\section{Introduction}

Competitive resource allocation describes scenarios in which multiple entities expend finite resources to acquire valuable assets, with applications ranging from advertising expenditures and military conflicts to multi-agent task allocation in engineering systems. A natural mathematical framework for studying such scenarios is contest games \cite{tullock,csf,survey:contest,Robson_2005,ContestTheory2017,Kovenock_handbook_2012}, in which players submit bids in a lottery-like mechanism and the probability of winning depends on their relative investments. A canonical model within this framework is the \emph{Tullock contest}, which offers tractable equilibrium analysis while capturing the fundamental tradeoff between resource expenditure and winning probability. Beyond simply optimizing one's own allocation, this framework enables the study of a wide range of strategic scenarios such as the formation of coalitions, asymmetric access to resources, or the subdivision of assets \cite{shah2024inefficient,paarporn2021division}. This literature has shed light on the variety of ways that one can further one's competitive position, which is of significant interest in applications across economics and engineering.

In this work, we study multi-player Tullock contests in which a central coordinator has a stake in the collective performance of a subset of players, referred to as subordinates, competing against independent opponents. The coordinator cannot directly dictate the subordinates' actions, but can influence their behavior by designing their individual objectives. The central question we address is: how should the coordinator design these objectives to best advance its own competitive position?


This question shares research goals with several bodies of literature. The field of utility design has focused on designing agent utility functions in multi-agent systems to minimize the performance gap between Nash equilibrium and optimal behavior; c.f., \cite{paccagnan2022utility} and references therein. In the context of traffic routing, a rich literature has studied taxation mechanisms that better align the incentives of self-interested users with social welfare \cite{beckmann1956studies,paccagnan2021optimal,meir2016marginal}. In the contests literature, several studies have examined group and individual-level incentive structures and their implications for inter-group competition \cite{munster2007simultaneous,konrad2009strategy}. A key distinction of our work from these bodies of literature is that the coordinator designs incentives for only a subset of the players, i.e., the subordinates, while the remaining players act as strategic opponents outside the coordinator's influence.

In this paper, the coordinator selects a mechanism that influences subordinate behavior by designing their individualized valuations of the contested prize. This particular choice of mechanism preserves the Tullock structure of the subordinates' objectives, enabling us to leverage our equilibrium characterization to address the design problem. We first establish equilibrium characterizations of the underlying multi-player Tullock contest, deriving closed-form expressions for equilibrium bids and payoffs as a function of each player's valuation and per-unit cost. Leveraging these results, we then establish our main contributions: characterizations of the optimal valuations that maximize the coordinator's objective. We derive explicit analytical expressions for the case of two subordinates competing against a single opponent, and for an arbitrary number of subordinates we establish a structural result that reduces the coordinator's optimization to a tractable two-variable problem. A key finding is that the optimal design systematically exploits cost asymmetry among subordinates, i.e., by assigning valuations that account for differences in per-unit costs, the coordinator can steer the equilibrium in its favor more effectively than any symmetric design would allow.

\section{Multi-player Tullock Contest}\label{sec:model}
Consider a set of players $\N = \{ 1,\cdots, n\}$ that compete over a single contest. In order to participate in the contest, each player must submit a bid $x_i \geq 0$ and the prize is awarded based on the collective bid $x \in \R_{\geq0}^n$.  The mechanism used to award the prize is a Tullock contest~\cite{tullock,csf} which takes the form, \begin{equation} \label{eq:ui}
    U_i( x ) = v_i \frac{x_i}{ \sum_{j\in\N} x_j } - c_i x_i,
\end{equation}
where the valuation $v_i>0$ characterize how profitable is for player $i$ to obtain the contested asset and the per-unit cost $c_i>0$ represent the efficiency of player $i$'s bid. 

For given values of $\{v_i\}_{i\in\N}$ and $\{c_i\}_{i\in\N}$ we model the emergent behavior as the Nash equilibrium of the utility functions defined in Equation~\eqref{eq:ui}. A Nash equilibrium is defined as the collective bids $x^*\in\R_{\geq0}^n$ that satisfy, \begin{equation*}
     U_i( x_i^*, x_{-i}^* ) \geq U_i( x_i, x_{-i}^* )  ~~\forall x_i \geq 0 \text{ and } i \in \N
\end{equation*}
where $x_{-i} = \prth{ x_1, \cdots, x_{i-1}, x_{i+1}, \cdots, x_n }$. This is, no single player has an incentive to unilaterally change their bid at equilibrium.  Note that this emergent behavior depends on the values of $v_i$ and $c_i$ for all $i\in\N$. That is, the value of $x^*$ is a function of $\{v_i\}_{i\in\N}$ and $\{c_i\}_{i\in\N}$. However, for simplicity, we omit the arguments for $x^*$.

\begin{figure}[hbt]
    \centering
    \includegraphics[width=0.23\textwidth]{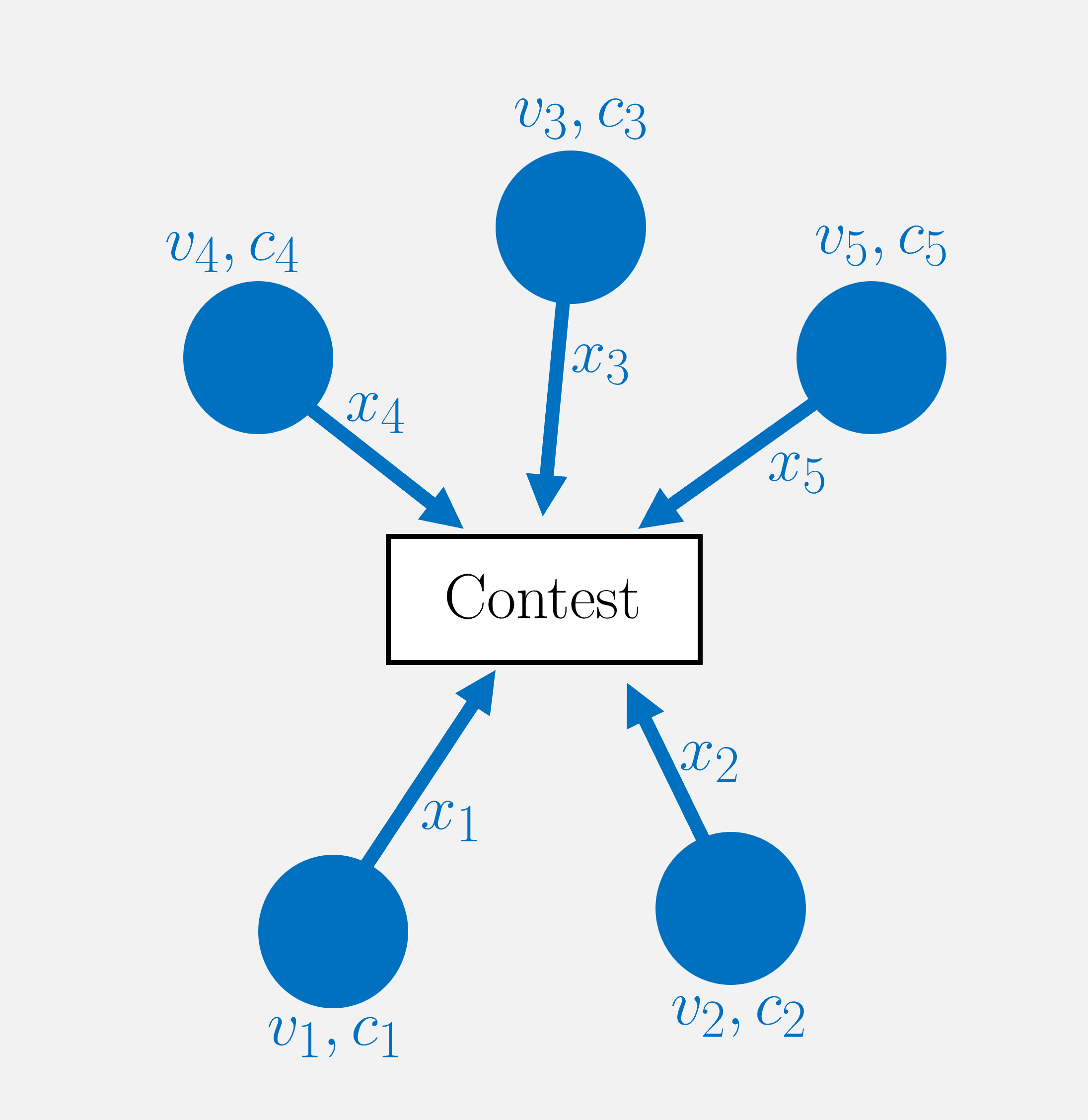}
    \caption{ Multi-player Tullock contest. In them, a group of players, with valuations $v_i$ and per-unit costs $c_i$, submit bids $x_i$ in order to obtain a contested prize.   }
    \label{fig:model1}
\end{figure}

\subsection{Nash Equilibrium Characterization for the Multi-player Tullock Contest}\label{sec:eq}
Given the model presented in Section~\ref{sec:model}, we present a characterization of the Nash equilibrium of the multi-player Tullock contest. This characterization give us some insight on how the parameters of the game affect the resulting behavior of the players.  With this in mind, Lemma~\ref{lm:neq} presents the equilibrium bids $x^*$ and their corresponding equilibrium payoffs $U_i^* := U_i(x^*)$ for arbitrary valuations $\{v_i\}_{i\in\N}$ and per-unit costs  $\{c_i\}_{i\in\N}$.
\begin{lemma} \label{lm:neq}
    For valuations $\{v_i\}_{i\in\N}$ and per-unit costs  $\{c_i\}_{i\in\N}$, the bids used at the Nash equilibrium are, \begin{equation} \label{eq:xopt}
        x_i^* = \alpha \sqr{ 1 - \frac{c_i}{v_i} \alpha }^+,
    \end{equation}
    where $[a]^+ = \max\{0,a\}$ and $\alpha$ is the unique solution to, \begin{equation} \label{eq:alpha}
        \sum_{i\in\N} \sqr{ 1 - \frac{c_i}{v_i} \alpha }^+ = 1.
    \end{equation}
    Moreover, their equilibrium payoff will be, \begin{equation} \label{eq:uopt}
         U_i^* = [ v_i - c_i \alpha ]^+.
    \end{equation}
\end{lemma}

An important distinction between $2$-players Tullock contest and multi-player Tullock contests is that the latest admit equilibrium bids equal to zero, i.e., $x_i^*=0$. This occurs because there exist scenarios where other players' equilibrium bids are high enough to prevent players to obtain a significant profit using strictly positive bids. The subset of players that set $x_i^*=0$ is intrinsically related to the value of $\alpha$ that solves Equation~\eqref{eq:alpha}. If $\alpha \geq \frac{v_i}{c_i}$ then player $i$ will prefer to not participate in the contest. In Remark~\ref{rmk:alpha} we present an algorithmic way of calculating the value of $\alpha$. 
\begin{remark} \label{rmk:alpha}
    Assume that the indices of the players are sorted such that $\frac{v_1}{c_1} \leq \frac{v_2}{c_2} \leq \cdots \leq \frac{v_n}{c_n}$. Then, \[ \alpha = (n-i^*) \prth{ \sum_{k=i^*}^n \frac{c_k}{v_k} }^{-1}, \] 
    where $i^*$ is the minimum index $i$ that satisfies, \[ (n-i) \frac{c_i}{v_i} \leq \sum_{k=i}^n \frac{c_k}{v_k} \leq (n-i) \frac{c_{i-1}}{v_{i-1}}, \]
    for $i\in\{2,\cdots,n-1\}$ or, when $i=1$, \[ (n-1) \frac{c_1}{v_1} \leq \sum_{k=1}^n \frac{c_k}{v_k}.  \]
    Additionally, for every $i < i^*$, we have that $x_i^* = 0$ and $U_i^* = 0$.
\end{remark}

Remark~\ref{rmk:alpha} reveals that the decision of abstaining to participate in the contest is related to the relative per-unit costs $\frac{c_i}{v_i}$. If the relative per-unit cost of player $i$ is significantly higher than the average of the smaller relative per-unit costs then, player $i$ will not participate in the contest. In Remark~\ref{rmk:alpha3} we present the particular solution for $n=3$. 
\begin{remark} \label{rmk:alpha3}
    For $n=3$, assume that $\frac{v_1}{c_1} \leq \frac{v_2}{c_2} \leq \frac{v_3}{c_3}$. Then, \begin{itemize}
        \item If $\frac{c_2}{v_2} + \frac{c_3}{v_3} \leq \frac{c_1}{v_1}$ then, $x_1^*=0$, \begin{align*}
            x_2^* = \frac{ {c_3}/{v_3} }{ \prth{ {c_2}/{v_2} + {c_3}/{v_3} }^2}, && \text{and} && x_3^*= \frac{ {c_2}/{v_2} }{\prth{ {c_2}/{v_2} + {c_3}/{v_3} }^2},
        \end{align*}
        and their respective payoffs are $U_1^* =0$, \begin{align*}
            U_2^* = v_2 \,\frac{ {c_3}/{v_3} }{ {c_2}/{v_2} + {c_3}/{v_3} }, && \text{and} && U_3^*= v_3\, \frac{ {c_2}/{v_2} }{ {c_2}/{v_2} + {c_3}/{v_3} }.
        \end{align*}

        \item If $\frac{c_1}{v_1} \leq \frac{c_2}{v_2} + \frac{c_3}{v_3}$ then, \[ x_i^* = \frac{ 2\prth{ {c_1}/{v_1} + {c_2}/{v_2} + {c_3}/{v_3} } - 4{c_i}/{v_i} }{\prth{ {c_1}/{v_1} + {c_2}/{v_2} + {c_3}/{v_3} }^2}, \]
        with equilibrium payoffs, \[ U_i^* = v_i - \frac{2c_i}{ {c_1}/{v_1} + {c_2}/{v_2} + {c_3}/{v_3} } \]
    \end{itemize} 
\end{remark}

Specifically, Remark~\ref{rmk:alpha3} states that if the highest relative per-unit cost $\frac{c_1}{v_1}$ is bigger than the sum of the other two relative per-unit costs  $\frac{c_2}{v_2} + \frac{c_3}{v_3}$ then, player $1$ will prefer to not participate on the Tullock contest. 



\begin{figure*}[!htb]
    \centering
    \subfloat[\label{fig:model2}]{\includegraphics[height=24ex]{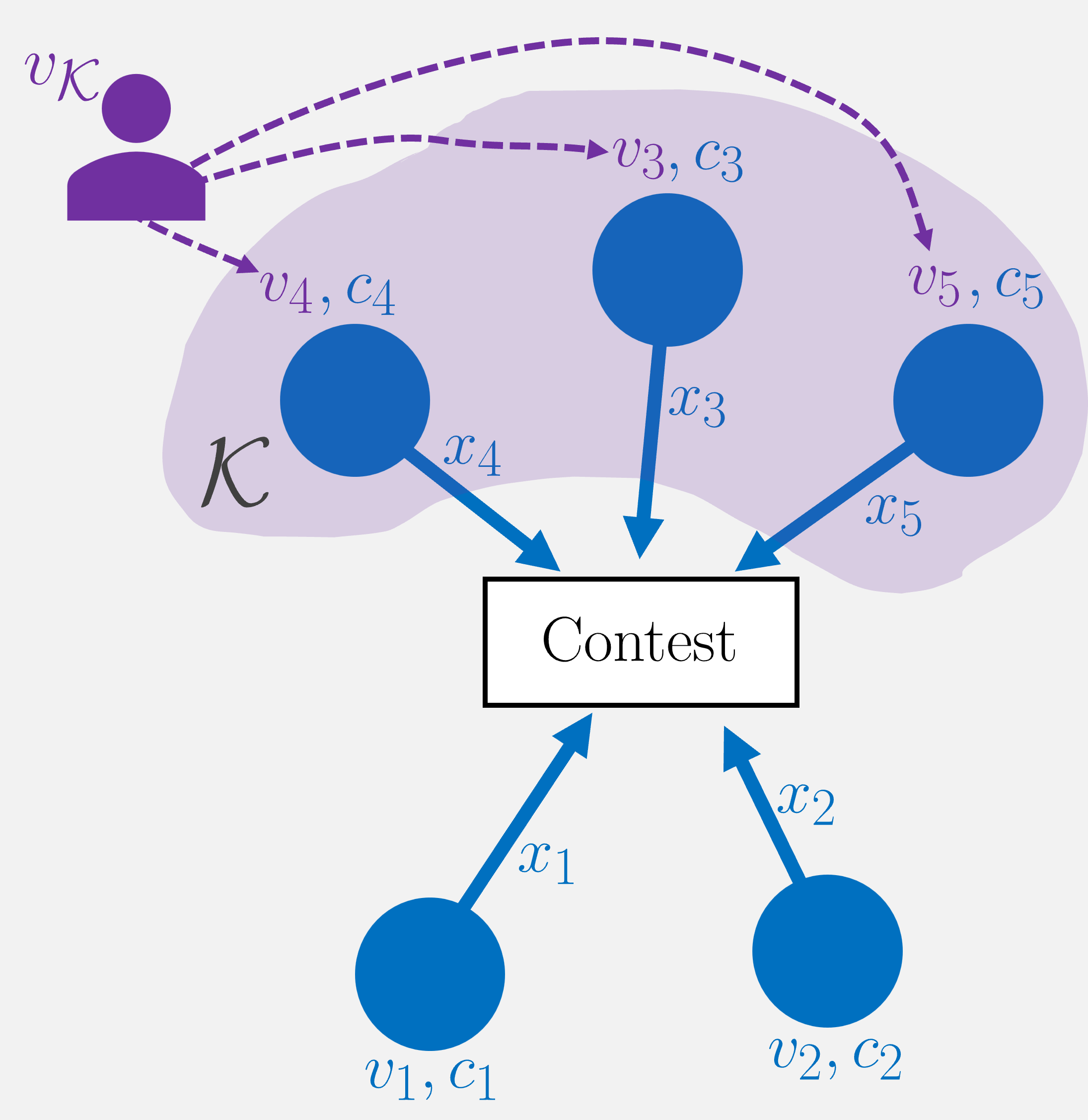}} \hfil
    \subfloat[\label{fig:ex1-1}]{\includegraphics[height=24ex]{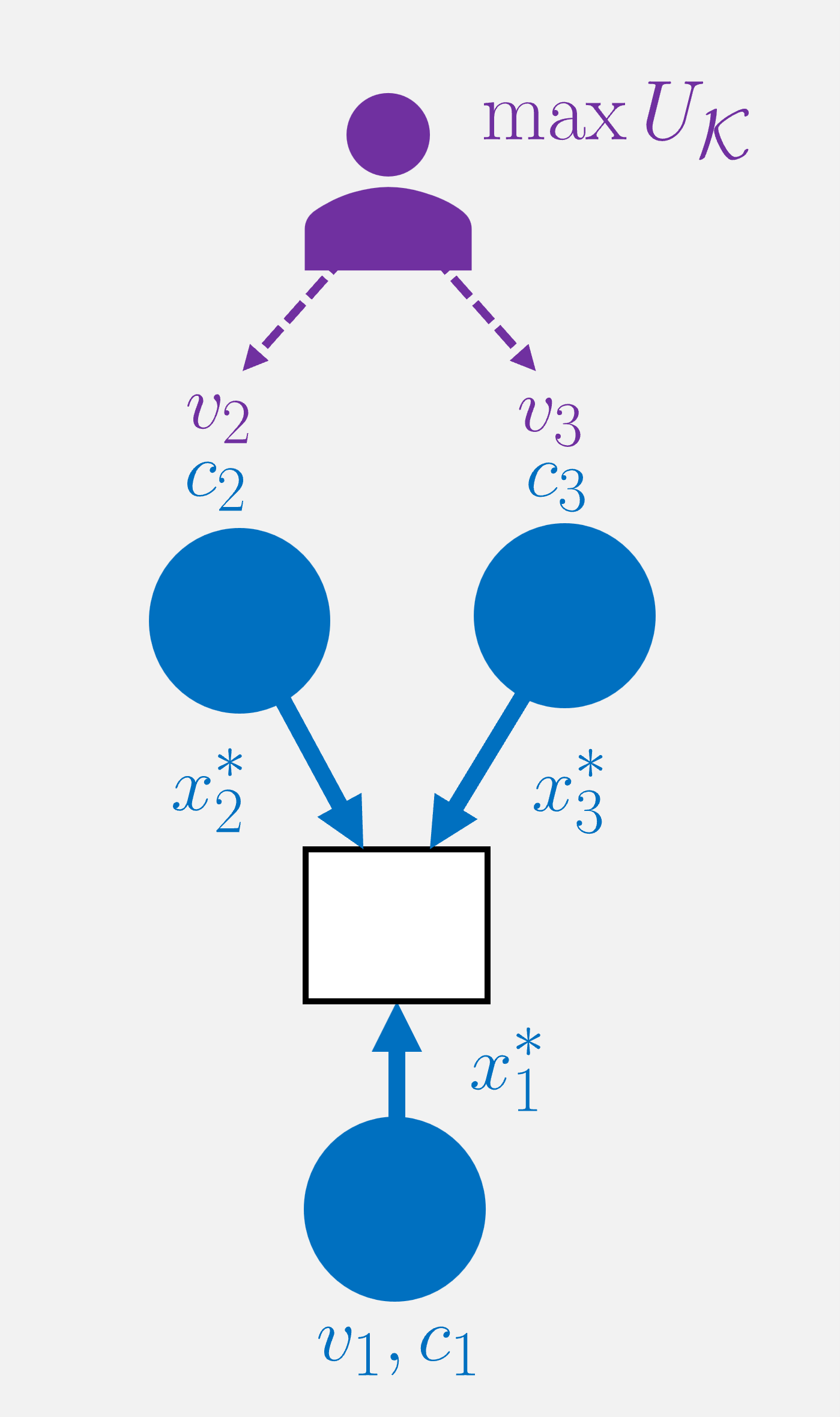}} \hfil
    \subfloat[\label{fig:ex1-2}]{\includegraphics[height=24ex]{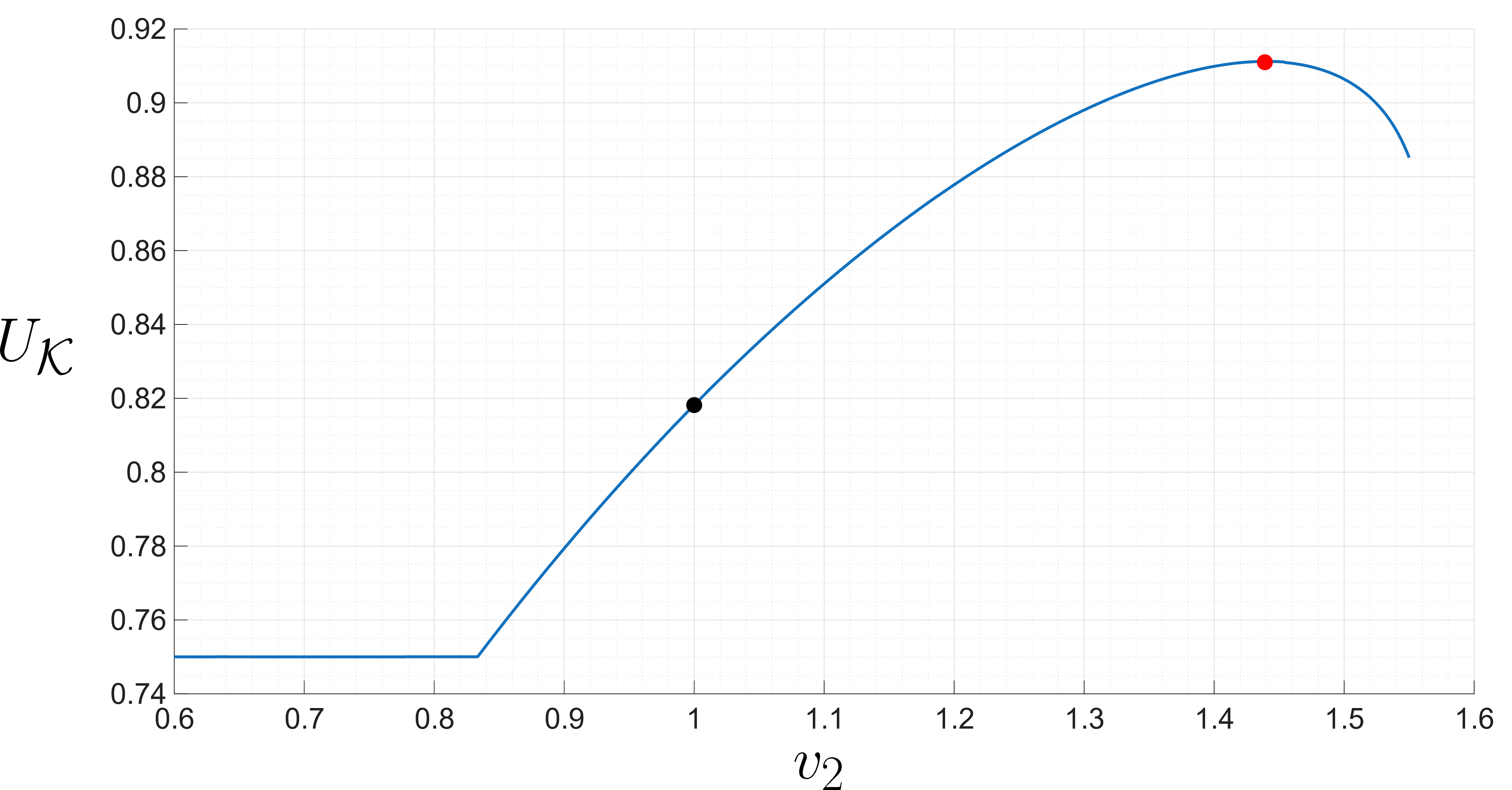}} 
    \caption{ (a) Multi-player Tullock contest with a central coordinator. Here, a central coordinator participates in the contest to obtain an item valued at $v_\K$. The coordinator participates in the contest is through the bids of its subordinate players $x_i$ for $i\in\K$. While the coordinator cannot directly determine the bids, it can influence the subordinate players by revealing their values $v_i$ for $i\in\K$. (b) For the case where $n=3$ and $\K = \{2,3\}$ the coordinator can decide the values $v_2$ and $v_3$ in order to steer the resulting equilibrium,  $x_i^*$ for $i\in\{1,2,3\}$, into a more desirable one. Note that for the case of $k=2$ the feasible valuations defined by Equation~\eqref{eq:const} are parametrized by a line segment. (c) By fixing $v_1 = v_\K = 1$, $c_1=9$, $c_2=10$ and $c_3 = 3$ we can parametrize the entire solution space using just $v_2$. We can observe then that the utility of the coordinator depends on the choice of $v_2$. By highlighting the baseline design of $v_2=v_3=v_\K$ (black circle) and the maximum attainable value (red circle), can be observe than there values of $v_2$ and $v_3$ can be optimally designed to improve the coordinator obtained profit, as in Equation~\eqref{eq:uk}. }
    \label{fig:ex:graph}
\end{figure*}

\section{Multi-player Tullock Contest with a Central Coordinator} \label{sec:model:coord}
Now that we understand how the game’s parameters influence the players’ emergent behavior, we investigate whether these parameters can be designed to steer that behavior toward more desirable outcomes. With that in mind, we introduce a central coordinator that also participates in the Tullock contest. We assume that the coordinator cares about the bids of a subset $\K\subsetneq \N$ of the bidding players with $|\K| = k < n$ and its benefit is derived from a contest between the subordinate players and the individual players. Thus, its utility can be written as, \begin{equation} \label{eq:uk}
    U_\K( x ) = v_\K \frac{\sum_{i\in\K} x_i }{ \sum_{j\in\N} x_j },
\end{equation}
where the valuation $v_\K >0$ characterize how valuable is the contested asset for the coordinator. Note that the coordinator participates in the contest only through the bids of the subordinate players, i.e., $x_i$ for $i\in\K$. However, the coordinator cannot directly specify the subordinate players' bids. Instead, their bids are guided through their local objectives. Therefore, the coordinator must incentivize the subordinate players to optimize its objective as in Equation~\eqref{eq:uk}. In this work, we assume that the mechanism that the coordinator has to influence the subordinate players' behavior is by redistributing its profit among them. Then, the utility for the subordinate players can be written as, \begin{equation*}
    U_i(x) = \bar{v}_i(x_\K) U_\K(x) - c_i x_i, ~~~ \forall i \in \K,
\end{equation*}
where $x_\K$ is the collective bid of the subordinate players and the function $\bar{v}_i(x_\K)$ is the distribution mechanism using by the coordinator. To ensure that the choice of $\{\bar{v}_i(\cdot)\}_{i\in\K}$ represents a valid mechanism we ensure that $\sum_{i\in\K} \bar{v}_i(x_\K) = 1$ for all $x_\K \in \R^k_{\geq 0}$. This is, the coordinator is redistributing completely its profit among its subordinate players. In our work, we focus on the redistribution mechanism defined as, \begin{equation} \label{eq:barvi}
    \bar{v}_i(x_\K) = v_i \frac{x_i}{ \sum_{j\in\K} x_j },
\end{equation}
where the coordinator is able to choose the values of $v_i$ for all $i\in\K$. An interpretation for this is that the coordinator can report how profitable is the contested asset to its sub-players. However, the capability for each player, represented by their per-unit cost $c_i$, is assumed to be intrinsic to them. For this particular choice, a valid incentive mechanism must satisfy, 
\begin{equation} \label{eq:const} 
    \sum_{i\in\K} v_i x_i = v_\K \sum_{i\in\K} x_i,
\end{equation}
for all possible bids $x_i$ with $i\in\K$. Therefore, we investigate if the central coordinator can report valuations that drives the behavior of the players, both subordinate and individual ones, into a more desirable outcomes in terms of its utility, as in Equation~\eqref{eq:uk}. This design problem can be written as the optimization problem, 
\begin{equation} \label{eq:opt:inf} \begin{aligned}
        \max_{\{ v_i\}_{i\in\K} } \quad    & v_\K \frac{\sum_{i\in\K} x_i^* }{ \sum_{j\in\N} x_j^* }  \\
        \text{subject to}             & ~~\{x_i^*\}_{i\in\N} \text{ is a Nash Equilibrium } \\
                                & ~~\{v_i\}_{i\in\N} \text{ is a valid mechanism }
    \end{aligned}
\end{equation}

Given the redistribution mechanism described in Equation~\eqref{eq:barvi}, the utilities for all players are of the form of Equation~\eqref{eq:ui}. However, for $i\in\K$, the coordinator will decide the values $v_i$ that the players will use. THerefore, we can leverage on the analysis of the Nash equilibrium for multi-player Tullock contests, as presented in Section~\ref{sec:eq}, to address the design question presented in Equation~\eqref{eq:opt:inf}.

\subsection{Incentive Design for $3$-player Tullock Contest}\label{sec:2v1}
Using the results from Section~\ref{sec:eq} we can address the question on how to optimally design the values of the subordinate players. In particular, we start with the smallest non-trivial scenario where the incentive design problem can be formulated: $n=3$ and $k=2$. Without losing generality, we assume that $\K=\{2,3\}$. Then, the incentive design problem can be written as the optimization problem,
\begin{equation}\label{eq:opt3} \begin{aligned}
        \max_{v_2,v_3} \quad    & v_\K \frac{x_2^* + x_3^*}{x_1^*+x_2^*+x_3^*}  \\
        \text{subject to }      & v_2 x_2^* + v_3 x_3^* = v_\K( x_2^* + x_3^* ) 
    \end{aligned}
\end{equation}
where the constraint comes from the condition for feasible incentive mechanisms in Equation~\eqref{eq:const}. Recall that the values of $\{x_i^*\}_{i\in\N}$ depend on the choice of $\{v_i\}_{i\in\N}$ and $\{c_i\}_{i\in\N}$. Therefore, we state the values of $v_2^*$ and $v_3^*$ that maximize the coordinator's profit in Theorem~\ref{thm:2v1}.
\begin{theorem} \label{thm:2v1}
    Given a valuation for the coordinator $v_\K$, valuation for the individual player $v_1$ and per-unit costs for all the players $c_1,c_2$ and $c_3$, the valuations that maximize the payoff of the coordinator, as in Equation~\eqref{eq:uk}, are, \begin{itemize}
        \item If $v_\K \frac{c_1}{v_1} \geq 2\sqrt{c_2 c_3}$ then, any solution that satisfy $c_2 (v_3^*)^2 + c_3(v_2^*)^2 = v_\K( c_2v^*_3 + c_3v^*_2 )$ and,
        \[ \frac{c_1}{v_1} \geq \frac{c_2}{v^*_2} + \frac{c_3}{v^*_3}, \] 
        is an optimal solution with $U_\K^* = v_\K$.
        
        \item If $v_\K \frac{c_1}{v_1} \leq 2\sqrt{c_2 c_3}$ then, 
        \[ v_i^* = \frac{2v_\K \frac{c_1}{v_1} + (\sqrt{c_2} - \sqrt{c_3})^2} {\frac{c_1}{v_1}( \sqrt{c_2} + \sqrt{c_3} )}\, \sqrt{c_i}, \]
        for $i\in\{2,3\}$ and, 
        \[ U_\K^* = v_\K \, \frac{2v_\K \frac{c_1}{v_1} + (\sqrt{c_2} - \sqrt{c_3})^2 }{ v_\K \frac{c_1}{v_1} + c_2 + c_3 }. \]
    \end{itemize}
\end{theorem} 

Theorem~\ref{thm:2v1} shows that the behavior of the optimally designed valuations $v_2$ and $v_3$ belong to two possible cases. First, when the relative per-unit cost of the individual player $\frac{c_1}{v_1}$ is bigger than the quantity $2\frac{\sqrt{c_2c_3}}{v_\K}$. In this case, the coordinator can pick values that force player $1$ to not compete in the contest, i.e., $x_1^*=0$. On the other hand, if player $1$ is able to participate, the coordinator picks valuations that are proportional to $\sqrt{c_i}$ for each $i\in\K$. This means that the coordinator gives a higher valuation to the player with the higher cost in order to increase their equilibrium bid. However, this comes with the cost of reducing the equilibrium bid of the player with the lowest cost. In other words, the coordinator is exploiting the differences between the costs $c_2$ and $c_3$ to obtain a more desirable outcome. Moreover, when $c_2 = c_3$, the optimal valuations are just $v_2^*=v_3^*=v_\K$ verifying that the smart design of the valuations comes from the asymmetry of the costs. 

\subsection{Incentive Design for Arbitrarily Large Multi-player Tullock Contests}\label{sec:results}
In this section, we extend the results presented in Section~\ref{sec:2v1}. Intuitively, we can just use Lemma~\ref{lm:neq} to find the equilibrium bids for an arbitrary number of players. However, as presented in Remark~\ref{rmk:alpha}, the solution of Equation~\eqref{eq:alpha} depends on the order of the relative per-unit costs $\frac{c_i}{v_i}$ for $i\in\N$. This means that we would need to solve an optimization problem for each possible order which suppose an explosion in the complexity of the problem. Instead, using the definitions in Equations~\eqref{eq:xopt}~and~\eqref{eq:alpha}, we can redefine the valuation design problem that the coordinator needs to solve. Thus, the optimization problem for the coordinator can be written as, 
\begin{equation}\label{eq:opt} \begin{aligned}
        \max_{\alpha\geq 0,\{ v_i\}_{i\in\K} } \quad    & \sum_{j\in\K} [v_j - c_j \alpha]^+  \\
        \text{subject to}             & \sum_{j\in\K} [v_j - c_j \alpha]^+ = v_\K \sum_{j\in\K} \sqr{ 1 - \frac{c_k}{v_j} \alpha}^+ \\
                                & \sum_{j\in\K} \sqr{ 1 - \frac{c_j}{v_j} \alpha}^+ + \sum_{j\notin\K} \sqr{ 1 - \frac{c_j}{v_j} \alpha}^+ = 1
    \end{aligned}
\end{equation}

In Equation~\eqref{eq:opt} we include $\alpha$ as one of the optimization variables while including Equation~\eqref{eq:alpha} as one of the constraints. This guarantee that the solution of $\alpha$ obtained solving Equation~\eqref{eq:opt} is the one corresponding to the equilibrium of the multi-player Tullock contest. Moreover, we can replace the values of $x_i^*$ to the expression in Equation~\eqref{eq:xopt}. While the complexity of finding explicit solution has not been reduced, Equation~\eqref{eq:opt} allow to state some structural properties about the optimal valuations $v_i^*$ for $i\in\K$. 
\begin{theorem} \label{thm:vi}
    Given a valuation for the coordinator $v_\K$, valuations for the individual players $\{v_i\}_{i\notin\K}$ and per-unit costs for all the players $\{c_i\}_{i\in\N}$, the valuations that maximize the payoff of the coordinator, as in Equation~\eqref{eq:uk}, satisfy, \begin{equation*}
        \frac{ \bar{v}_i^* }{ \sqrt{c_i} } = \frac{ \bar{v}_j^* }{ \sqrt{c_j} }
    \end{equation*}
    for all $i,j \in \K$. Equivalently, the optimal valuations can be written as $\bar{v}_i^* = \beta \sqrt{c_i}$ with some $\beta \geq0$ for all $i \in\K$.
\end{theorem} 

Theorem~\ref{thm:vi} asserts that the optimal valuations satisfy $v_i^* = \beta \sqrt{c_i}$. We can compare this with our result in Theorem~\ref{thm:2v1}. If $v_\K \frac{c_1}{v_1} \leq 2\sqrt{c_2c_3}$ then, we have that $\beta = \frac{2v_\K \frac{c_1}{v_1} + (\sqrt{c_2} - \sqrt{c_3})^2} {\frac{c_1}{v_1}( \sqrt{c_2} + \sqrt{c_3} )}$. Or, if we have that $v_\K \frac{c_1}{v_1} \geq 2\sqrt{c_2c_3}$, picking $\beta = \frac{v_\K(\sqrt{c_2}+\sqrt{c_3})}{2\sqrt{c_2c_3}}$ is a feasible solution. However, with Theorem~\ref{thm:vi}, we can extend this behavior into arbitrarily large multi-player Tullock contests. Moreover, if we constrain the optimal valuations to the ones that satisfy $v_i^*= \beta\sqrt{c_i}$ then, we can drastically reduce the complexity of the optimization problem that the coordinator has to solve, as stated in Remark~\ref{rmk:opt}.
\begin{remark} \label{rmk:opt}
    The optimal valuations can be computed by solving the following reduced optimization problem, 
    \begin{equation}\label{eq:max2} \begin{aligned}
        \max_{\alpha \geq 0,\beta \geq 0} &~ \sum_{k\in\K} \sqrt{c_k}[ \beta - \alpha \sqrt{c_k} ]^+  \\
        \text{subject to} 
        & \sum_{k\in\K} [\beta - \alpha \sqrt{c_k}]^+ + \beta \sum_{k\notin\K} \sqr{1-\frac{c_k}{v_k}}^+ = \beta \\
        & \sum_{k\in\K} \prth{ \beta \sqrt{c_k} - v_\K }[\beta-\alpha \sqrt{c_k} ]^+ = 0
    \end{aligned} 
    \end{equation}
\end{remark}

Equation~\eqref{eq:max2} in Remark~\ref{rmk:opt} comes from replacing $v_i$ with $\beta\sqrt{c_i}$ in Equation~\eqref{eq:opt}. However, it is important to note that the number of variables got reduced from $k+1$ to $2$, significantly reducing the dimension of the optimization problem that the coordinator requires to solve.

\section{Conclusion}\label{sec:conclusion}
The contributions of this work provide insight into how the Nash equilibrium of a contest game can be steered by modifying the information revealed to its participants. Using multi-player Tullock contest as a modeling framework, we characterize how a central coordinator can indirectly influence the outcome of the game by misreporting the subordinate players’ valuation of the contested item.  These results offer an initial analytical tool for understanding the role of information in competitive resource allocation scenarios. Future work includes analyzing more complex information structures, extending the model to incorporate multiple coordinators, and including uncertainty in other game parameters among different players.  

\bibliographystyle{IEEEtran}
\bibliography{bib,kp_sources}

\appendix

\subsection{Proof of Lemma~\ref{lm:neq} }
Here, the utility for all players have the form of Equation~\eqref{eq:ui}. Then, the first order conditions state that, \begin{equation*} \begin{aligned}
    v_i \frac{ \sum_{k\neq i} x_k^* }{ (\sum_{k\in\N} x_k^* )^2 } - c_i = 0,
\end{aligned} \end{equation*}
for all $i\in\N$. Let us define $\alpha = \sum_{i\in\N} x_i^* > 0$. Then, the above condition can be rewritten as, \begin{equation*} \begin{aligned}
        v_i \frac{\alpha - x_i^*}{\alpha^2} = c_i  \iff x_i^* = \alpha \prth{ 1 - \frac{c_i}{v_i} \alpha }. 
\end{aligned} \end{equation*} 
Since we have to guarantee that $x_i^* \geq 0$ then, \begin{equation*}
    x_i^* = \alpha \sqr{ 1 - \frac{c_i}{v_i} \alpha }^+,
\end{equation*}
which depends on the value of $\alpha$. To find it note that, \begin{equation*} \begin{aligned}
    \alpha = \sum_{i\in\N} x_i^* = \sum_{i\in\N} \alpha \sqr{ 1 - \frac{c_i}{v_i} \alpha }^+ 
    \iff  \sum_{i\in\N} \sqr{ 1 - \frac{c_i}{v_i} \alpha }^+ = 1.
\end{aligned} \end{equation*}
Now, the function, $f(\alpha) = \sum_{i\in\N} \sqr{ 1 - \frac{c_i}{v_i} \alpha }^+$ is strictly decreasing with $f(0)=n$. Therefore, there is an unique value of $\alpha$ that satisfy $f(\alpha)=1$. Now, for the utilities at equilibrium note that, 
\begin{equation*} \begin{aligned}
    U_i^* = v_i \frac{x_i^*}{ \sum_{k\in\N} x_k^* } = \frac{v_i}{\alpha} \alpha \sqr{ 1 - \frac{c_i}{v_i} \alpha }^+ = [v_i - c_i \alpha]^+.
\end{aligned} \end{equation*}

\subsection{Proof of Remark~\ref{rmk:alpha} }
If we assume that the indices of the players are sorted such that, $\frac{v_1}{c_1} \leq \frac{v_2}{c_2} \leq \cdots \leq \frac{v_n}{c_n}$ then, we have two possibilities. First, if $\alpha \leq \frac{v_1}{c_1}$ then, \begin{align*}
    &\sum_{i\in\N} \sqr{ 1 - \frac{c_i}{v_i} \alpha }^+ = \sum_{i\in\N} \prth{ 1 - \frac{c_i}{v_i} \alpha } = n - \alpha \prth{\sum_{i\in\N} \frac{c_i}{v_i} } = 1.\\
    &\implies \alpha = (n-1)\prth{ \sum_{i\in\N} \frac{c_i}{v_i}  }^{-1}.
\end{align*}
Now, we need to guarantee that, \begin{align*}
    \alpha = (n-1)\prth{ \sum_{i\in\N} \frac{c_i}{v_i}  }^{-1} \leq \frac{v_1}{c_1} \iff (n-1) \frac{c_1}{v_1} \leq \sum_{i\in\N} \frac{c_i}{v_i}
\end{align*}

If we assume instead that $\frac{v_{i-1}}{c_{i-1}} \leq \alpha \leq \frac{v_i}{c_i}$ for $i\in\{2,\cdots,n-1\}$ then,
\begin{align*}
    &\sum_{i\in\N} \sqr{ 1 - \frac{c_i}{v_i} \alpha }^+ = \sum_{k=i}^n \prth{ 1 - \frac{c_k}{v_k} \alpha } = 1 \\
    &\implies \alpha = (n-i)\prth{ \sum_{k=i}^n \frac{c_k}{v_k}  }^{-1}.
\end{align*}
Now, we need to guarantee that, \begin{align*}
    &\frac{v_{i-1}}{c_{i-1}} \leq (n-i)\prth{ \sum_{k=i}^n \frac{c_k}{v_k}  }^{-1} \leq \frac{v_i}{c_i} \\
    &\iff (n-i) \frac{c_{i-1}}{v_{i-1}} \geq \sum_{k=i}^n \frac{c_k}{v_k} \geq (n-i) \frac{c_i}{v_i}
\end{align*}
Thus, we can find the value of $\alpha$ by checking for the index $i^*$ that satisfy the inequalities. In addition, for all the indices $i<i^*$ we have that, \begin{align*}
    \sqr{ 1 - \frac{c_i}{v_i} \alpha }^+ = 0 \implies x_i^* = 0 \implies U_i^* = 0.
\end{align*}

\subsection{Proof of Theorem~\ref{thm:2v1} }
For the optimization problem in Equation~\eqref{eq:opt3} there are four possible cases, given Remark~\ref{rmk:alpha3}. For simplicity, let us define $w_i = \frac{c_i}{v_i}$ for $i\in\N$. 

\textbf{Case \#1:} $w_1 \geq w_2 + w_3$. Here, the players $2$ and $3$ can force player $1$ to set $x_1^*=0$. However, they still need to satisfy the constraint in Equation~\eqref{eq:opt3}. This can be written as, \begin{equation*}
    c_2 w_3^2 + c_3 w_2^2 = v_\K w_2 w_3 ( w_2 + w_3 ).
\end{equation*}
Together with the necessary constraint for this case, \begin{equation*}
    \frac{c_2 w_3^2 + c_3 w_2^2}{ v_\K w_2 w_3 } = w_2 + w_3 \leq w_1.
\end{equation*}
By analyzing when the previous equation is satisfied with equality we have, 
\begin{equation*} \begin{aligned}
    & c_3 \prth{ \frac{w_2}{w_3} }^2 - v_\K w_1 \prth{ \frac{w_2}{w_3} } + c_2 = 0, \\
    &\iff \frac{w_2}{w_3} = \gamma_{1,2} = \frac{v_\K w_1 \pm \sqrt{ (v_\K w_1)^2 -4c_2c_3}}{2c_3},
\end{aligned} \end{equation*}
which have real solutions only when $(v_\K w_1)^2 \geq 4c_2c_3$. 

\textbf{Case \#2:} $w_2 \geq w_1 + w_3$. Here, player $2$ is forced to set $x_2^* = 0$. Therefore, the constraint in Equation~\eqref{eq:opt3} make that $w_3 = v_\K$ and, consequently, the payoff obtained by the coordinator is equal to, $v_\K \frac{ {c_1}/{v_1} }{ {c_1}/{v_1} + {c_3}/{v_\K} }$. It is important to note that, since $x_2^*=0$, the coordinator can choose an arbitrary value for $w_2$ without violating the constraint. However, for a $w_2$ sufficiently large, it can avoid to be in the region defined for this case. 

\textbf{Case \#3:} $w_3 \geq w_1 + w_2$.  Similar to case \#2, in this case we have that $x_3^*=0$. Then, we obtain that $w_2 = v_\K$ and,$v_\K \frac{ {c_1}/{v_1} }{ {c_1}/{v_1} + {c_2}/{v_\K} }$. Also, as in case \#2, the coordinator can pick an arbitrary $w_3$.

\textbf{Case \#4:} $w_1 \leq w_2 + w_3$ and $w_2 \leq w_1 + w_3$ and $w_3 \leq w_1 + w_2$. In this case we have that, $x_i^* = \frac{2(w_1 + w_2 + w_3) - 4w_i}{(w_1 + w_2 + w_3)^2}$, which allow us to write Equation~\eqref{eq:opt3} as, 
\begin{equation*} \begin{aligned}
        &\max_{w_2,w_3} \quad  \frac{ 2v_\K w_1}{w_1 + w_2 + w_3}  \\
        \text{s.t. }      & \prth{ \frac{c_2}{w_2} + \frac{c_3}{w_3} }(w_1 + w_2 + w_3) = 2( v_\K w_1 + c_2 + c_3 ) 
    \end{aligned}
\end{equation*}

Equivalently, we can find the solution to, \begin{equation*} \begin{aligned}
        &\min_{w_2,w_3} \quad  w_2 + w_3 \\
        \text{s.t. }      & \prth{ \frac{c_2}{w_2} + \frac{c_3}{w_3} }(w_1 + w_2 + w_3) = 2( v_\K w_1 + c_2 + c_3 ) 
    \end{aligned}
\end{equation*}

 First-order conditions for this problem are, \begin{equation*}
     \frac{c_2}{ (w_2^*)^2 } = ( w_1 + w_2^* +w_3^* )^{-1} \prth{ \frac{1}{\lambda} + \frac{c_2}{w_2^* } + \frac{c_3}{w_3^* } } = \frac{c_3}{ (w_3^*)^2 }
 \end{equation*}
 where $\lambda$ is the Lagrange multiplier for the constraint. Thus, we obtain that $w_2^* = \sqrt{ \lambda w_1 c_2}$ and $w_3^* = \sqrt{ \lambda w_1 c_3}$. After replacing back into the constraint we obtain, \begin{equation*}
     \sqrt{\lambda} = \sqrt{w_1} \frac{\sqrt{c_2} + \sqrt{c_3} }{ 2v_\K w_1 + (\sqrt{c_2} - \sqrt{c_3})^2 }.
 \end{equation*}

 Therefore, \begin{equation*}
     w_i = \sqrt{ \lambda w_1 c_i} = w_1  \frac{\sqrt{c_2} + \sqrt{c_3} }{ 2v_\K w_1 + (\sqrt{c_2} - \sqrt{c_3})^2 } \sqrt{c_i},
 \end{equation*}
for $i\in\{2,3\}$ and, consequently, the utility for the coordinator will be, \begin{equation*}
    v_\K \frac{2v_\K w_1 + (\sqrt{c_2} - \sqrt{c_3} )^2 }{ v_\K w_1 + c_2 + c_3 }.
\end{equation*}

Now, if $(v_\K w_1)^2 \geq 2c_2c_3$ then, we can pick values of $\bar{v}_2$ and $\bar{v}_3$ that guarantee that $x_1^*=0$ i.e., the coordinator guarantees to win the contest. However, if $(v_\K w_1)^2 \leq 2c_2c_3$ then, the coordinator would have to pick between cases \#2, \#3 and \#4. Without losing generality, let us assume that $c_2 \leq c_3$. Therefore, \begin{equation*}
    v_\K \frac{ {c_1}/{v_1} }{ {c_1}/{v_1} + {c_3}/{v_\K} } \leq v_\K \frac{ {c_1}/{v_1} }{ {c_1}/{v_1} + {c_2}/{v_\K} },
\end{equation*}
this is, case \#3 is always preferred to case \#2. Now, we need to check if case \#4 is preferred to case \#3 or, equivalently, \begin{equation*} \begin{aligned}
            & v_\K \frac{2v_\K w_1 + (\sqrt{c_2} - \sqrt{c_3} )^2 }{ v_\K w_1 + c_2 + c_3 } \geq v_\K \frac{w_1}{w_1 + \frac{c_2}{v_\K}}, \\
    \iff    & \sqr{ w_1 + \frac{(\sqrt{c_2} - \sqrt{c_3})^2 }{v_\K} } \prth{ w_1 + \frac{c_2}{v_\K} } \geq w_1 \frac{c_3}{v_\K}, \\
    \iff    & v_\K w_1 + \prth{ 1 + \frac{c_2}{v_\K w_1} }( \sqrt{c_2} - \sqrt{c_3} )^2 \geq c_3 - c_2, \\
    \iff    & \sqr{ v_\K w_1 + \sqrt{c_2}(\sqrt{c_2} - \sqrt{c_3}) }^2 \geq 0.
\end{aligned} \end{equation*}

Then, if $(v_\K w_1)^2 \leq 2c_2c_3$, the coordinator will always choose to be in case \#4. 

\subsection{Proof of Theorem~\ref{thm:vi} }
First order conditions for Equation~\eqref{eq:opt} can be written as, \begin{equation*} \begin{aligned}
    (1-\lambda_1) \H( v_i^* - c_i \alpha ) - \H\prth{ 1-\frac{c_i}{v_i^*}\alpha } \frac{c_i \alpha}{ (v_i^*)^2 } (\lambda_1 + \lambda_2) = 0,
\end{aligned} \end{equation*}
for all $i\in\K$ where $\lambda_1$ and $\lambda_2$ are the Lagrange multipliers for the constraints in Equation~\eqref{eq:opt} and $\H(x)$ is the Heaviside step function. Note that $\H( v_i^* - c_i \alpha ) = \H\prth{ 1-\frac{c_i}{v_i^*}\alpha }$ then the above conditions can be written as, \begin{equation*} \begin{aligned}
    \H(v_i^* - c_i \alpha ) \sqr{ (v_i^*)^2 (1-\lambda_1)  - c_i \alpha (\lambda_1 + \lambda_2) } =0.
\end{aligned} \end{equation*}
Then, if $v_i^* \leq c_i\alpha$ then, the value of $v_i^*$ can be chosen arbitrarily because the corresponding $x_i^*$ will be $0$ anyways. And, if $v_i^* > c_i\alpha$ then, \begin{equation*} \begin{aligned}
     & (v_i^*)^2 (1-\lambda_1)  - c_i \alpha (\lambda_1 + \lambda_2) = 0 \iff \frac{(v_i^*)^2}{ c_i } = \frac{\alpha (\lambda_1 + \lambda_2)}{1-\lambda_1}.
\end{aligned} \end{equation*}

Since the value of $\frac{\alpha (\lambda_1 + \lambda_2)}{1-\lambda_1}$ does not depend on the index of the player $i$ then we can set, \begin{equation*}
     \frac{ v_i^* }{ \sqrt{c_i} } = \beta = \frac{ v_j^* }{ \sqrt{c_j} },
\end{equation*}
for any $i,j\in\K$. Or equivalently, $v_i^* = \beta \sqrt{c_i}$ for all $i\in\K$ and some $\beta>0$.


\end{document}